\documentclass[12pt]{article}
\usepackage{graphicx}

\newcommand\pubnumber{CIPANP2018-Myslik}
\newcommand\pubdate{\today}

\def\lbnl{Nuclear Science Division\\
Lawrence Berkeley National Laboratory, Berkeley, CA, USA}

\def\Title#1{\begin{center} {\Large #1 } \end{center}}
\def\Author#1{\begin{center}{ \sc #1} \end{center}}
\def\Address#1{\begin{center}{ \it #1} \end{center}}

\newcommand\pubblock{\rightline{\begin{tabular}{l} \pubnumber\\
         \pubdate  \end{tabular}}}
\newenvironment{Abstract}{\begin{quotation}  }{\end{quotation}}
\newenvironment{Presented}{\begin{quotation} \begin{center} 
             PRESENTED AT\end{center}\bigskip 
      \begin{center}\begin{large}}{\end{large}\end{center} \end{quotation}}

\def\beq{\begin{equation}}
\def\eeq#1{\label{#1}\end{equation}}
\def\eeqn{\end{equation}}

\def\beqa{\begin{eqnarray}}
\def\eeqa#1{\label{#1}\end{eqnarray}}
\def\eeqan{\end{eqnarray}}

\let\bar=\overbar

\def\Dslash{\not{\hbox{\kern-4pt $D$}}}
\def\dslash{\not{\hbox{\kern-2pt $\del$}}}

\def\msb{{\bar{\ssstyle M \kern -1pt S}}}

\begin{document}
\begin{titlepage}
\pubblock

\vfill
\Title{LEGEND: The Large Enriched Germanium Experiment for Neutrinoless
    Double-Beta Decay}
\vfill
\Author{Jordan Myslik \\ for the LEGEND Collaboration}
\Address{\lbnl}
\vfill
\begin{Abstract}
The lepton number violating process of neutrinoless double-beta decay could
    result from the physics beyond the Standard Model needed to generate the
    neutrino masses. Taking different approaches, the current generation of
    $^{76}$Ge experiments, the \textsc{Majorana Demonstrator} and GERDA, 
	lead the field in
    both the ultra-low background and energy resolution achieved. The next
    generation of neutrinoless double-beta decay experiments requires increased
    mass and further reduction of backgrounds to maximize discovery potential.
    Building on the successes of the \textsc{\mbox{Majorana} Demonstrator} and GERDA, the
    \mbox{LEGEND} collaboration has been formed to pursue a tonne-scale $^{76}$Ge
    experiment, with discovery potential at a half-life beyond $10^{28}$ years. The
    collaboration aims to develop a phased neutrinoless double-beta decay
    experimental program, starting with a 200 kg measurement using the existing
    GERDA cryostat at LNGS. These proceedings discuss the plans and physics reach of
    \mbox{LEGEND}, and the combination of R\&D efforts and existing resources being
    employed to expedite physics results.

\end{Abstract}
\vfill
\begin{Presented}
Thirteenth Conference on the Intersections of Particle and Nuclear Physics\\
Palm Springs, CA, USA,  May 29--June 3, 2018
\end{Presented}
\vfill
\end{titlepage}
\def\thefootnote{\fnsymbol{footnote}}
\setcounter{footnote}{0}
\section{Introduction}
Two neutrino double-beta decay (``$2\nu\beta\beta$-decay''), $(A,Z)\rightarrow(A,Z+2) + 2e^{-} +
2\bar{\nu}_{e}$, is allowed in the Standard Model, and observed in isotopes
e.g. $^{76}$Ge, $^{130}$Te, and $^{136}$Xe.  Neutrinoless double-beta decay
(``$0\nu\beta\beta$-decay''), $(A,Z)\rightarrow(A,Z+2) + 2e^{-}$, violates
lepton number conservation by two units, independently of any model, and if
observed, would be a clear signature of New Physics.  Such a non-standard
process is a key feature of multiple neutrino mass generation models.  The experimental 
signature of $0\nu\beta\beta$-decay is a peak at the Q-value of an isotope's
double-beta decay spectrum.  Among various isotopes, $^{76}$Ge-based
experiments have a high discovery potential.

Specifically in the light neutrino exchange model \cite{Rodejohann:2011mu},
there is a relationship (through the isotope's nuclear matrix element and axial
vector coupling constant) between the $0\nu\beta\beta$-decay half-life, the neutino mass
ordering, and the lightest neutrino mass. In this model, a $^{76}$Ge
experiment sensitive to T$_{1/2} > 10^{28}$~y will be able to probe the entire
inverted ordering (IO) scenario, and a significant fraction of the
normal ordering (NO) parameter space \cite{Agostini:2017jim}. This half-life is
therefore a reasonable target for $^{76}$Ge experiments.

Neutrinoless double-beta decay at a half-life of $10^{28}$~years would produce a signal of
$\sim 0.5$~counts/t$\cdot$y, demonstrating the need for ultra-low
backgrounds, a large fiducial mass, and a long counting time for an
experiment pursuing this goal, as can be seen in Figure~\ref{fig:dp-plot}.  To have $3\sigma$ discovery
potential at the bottom of the IO-allowed region with 10~ton$\cdot$yr of exposure, 
it would require a tonne-scale detector enriched to 88\% $^{76}$Ge with backgrounds less than 
$0.1$~counts/FWHM$\cdot$t$\cdot$y.

Currently there are two operating $^{76}$Ge $0\nu\beta\beta$-decay
experiments: the \textsc{Majorana Demonstrator} \cite{Abgrall:2013rze} and GERDA (the GERmanium
Detector Array) \cite{Ackermann:2012xja}.  Both experiments include germanium crystal detectors 
enriched to $\sim88$\% in $^{76}$Ge (detectors totalling 29.7~kg for the
\textsc{Majorana Demonstrator}, and 37.6~kg for GERDA).  Both experiments are
also located underground to shield against cosmic ray muon backgrounds: the \textsc{Majorana Demonstrator}
in the Sanford Underground Research Facility (SURF) in Lead, South
Dakota, USA, and GERDA at the Laboratori Nazionali del Gran Sasso (LNGS),
located between L'Aquila and Teramo in Italy.  To achieve its background goals,
the \textsc{Majorana Demonstrator} relied on placing the detectors in 2 vacuum 
cryostats surrounded by a traditional passive shield, with components made of ultra-clean
materials. GERDA took the novel approach of submerging its detectors in liquid
argon, which acts as an active shield, as backgrounds are tagged using their
resultant scintillation light.  Both experiments have lower backgrounds than
all $0\nu\beta\beta$-decay searches using other isotopes, with GERDA holding the
record of $\sim3$~counts/(FWHM$\cdot$t$\cdot$y) and a sensitivity of
$1.1\times10^{26}$~y at the 90\% confidence level \cite{zsigmond2018}, and have
better energy resolution than all $0\nu\beta\beta$-decay searches using other
isotopes, with the \textsc{Majorana Demonstrator} holding the record
of 2.5~keV~FWHM at the Q-value of 2039~keV.

\section{LEGEND}
The LEGEND (Large Enriched Germanium Experiment for Neutrinoless $\beta\beta$
Decay) collaboration plans to  develop a $^{76}$Ge-based double-beta decay
experimental program, building upon the successes of the \textsc{Majorana
Demonstrator} and GERDA. \mbox{LEGEND} will proceed in phases towards a neutrinoless
double-beta decay discovery potential at a half-life beyond $10^{28}$ years.
The best technologies will be selected based on lessons learned in GERDA and
the \textsc{Majorana Demonstrator}, as well as contributions from other groups.
To expedite physics results, LEGEND will use existing resources as appropriate.

\begin{figure}[!h]

    \includegraphics[width=\textwidth]{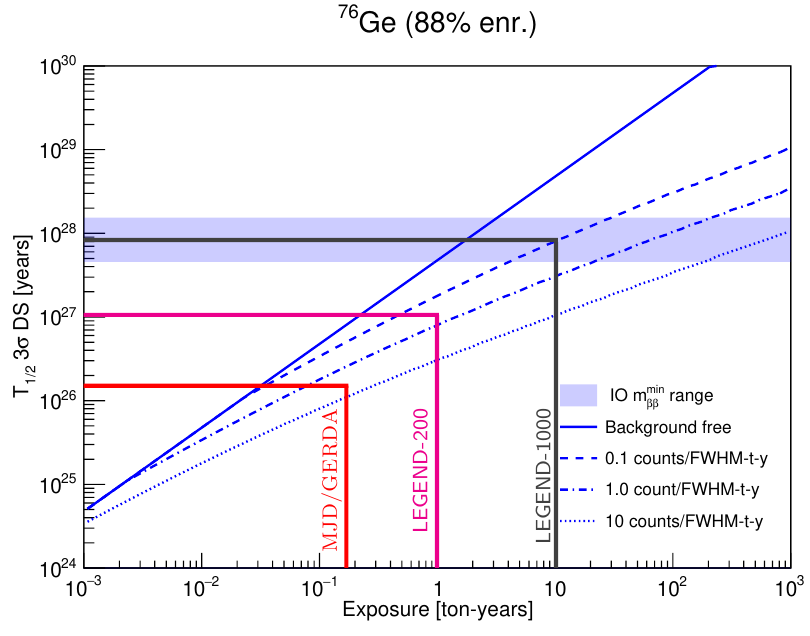}
    \caption{Neutrinoless double-beta decay half-life $3\sigma$ discovery
    potential as a function of exposure and background rate for detectors enriched
    to 88\% in $^{76}$Ge.  The bottom of the inverted ordering mass region
    ($m_{\beta\beta} = 17$~meV) is shown for unquenched axial vector coupling
    and nuclear matrix elements ranging from 3.5 to 5.5. The discovery
    potentials for the projected combined
    total exposures of the \textsc{\mbox{Majorana} \mbox{Demonstrator}} and GERDA are shown,
    along with the exposures of 1~ton$\cdot$y of LEGEND-200 at a background
    rate of 0.6~counts/FWHM$\cdot$t$\cdot$y, and
    10~ton$\cdot$y of LEGEND-1000 at a background rate of
    0.1~counts/FWHM$\cdot$t$\cdot$y. \label{fig:dp-plot}}

\end{figure}

In its first phase, LEGEND-200, the existing GERDA infrastructure at LNGS will
be modified, and up to 200~kg of detectors will be deployed in the cryostat.  This
initial phase permits early science with a world-leading experiment, helping to
maintain skilled workers on the timeline leading up to LEGEND-1000. LEGEND-200
has a background goal of less than $0.6$~counts/(FWHM$\cdot$t$\cdot$y).  
This required reduction in background has already been demonstrated as feasible 
in the \textsc{Majorana \mbox{Demonstrator}}, GERDA, and dedicated test stands.  
Achieving this background
rate will allow LEGEND-200 to reach a sensitivity greater than $10^{27}$~y
after accumulating 1~ton$\cdot$y of exposure.  The corresponding discovery
potential is shown in Figure~\ref{fig:dp-plot}.  LEGEND-200 physics data collection 
is expected to begin in 2021.

Subsequent stages will occur in new infrastructure, with 1000~kg of detectors 
deployed in stages up to the complete LEGEND-1000.  The background goal for
LEGEND-1000 is less than $0.1$~counts/(FWHM$\cdot$t$\cdot$y).  This background
reduction is necessary for LEGEND-1000 to achieve a $0\nu\beta\beta$-decay
discovery potential at a half-life greater than $10^{28}$ years on a reasonable
timescale (see Figure~\ref{fig:dp-plot}). The required depth to
keep cosmogenic activation backgrounds (e.g. $^{77\mathrm{m}}$Ge) within the background
budget is currently under investigation (e.g. \cite{Wiesinger:2018qxt}), 
and will be a contributing factor in the choice of site.  
The timeline of LEGEND-1000 is connected to the U.S. Department of Energy
down-select process for the next generation neutrinoless double-beta decay
experiment.

\section{Background reduction plans}
Multiple techniques are planned to achieve the background reduction required for LEGEND-200, 
and the further reduction required for LEGEND-1000.  
This section discusses a selection of these techniques.

\subsection{Electroformed copper}
Amongst the ultra-clean materials employed by the $\textsc{Majorana
Demonstrator}$ is electroformed copper.  Its uranium and thorium decay chain 
backgrounds each average $\le0.1$~$\mu$Bq/kg, which are lower than the majority
of the commercial OFHC copper used in the \textsc{Majorana Demonstrator} by at
least an order of magnitude \cite{Abgrall:2016cct}.  It is employed in the parts closest to
the detectors (e.g. detector mounts, inner copper shield) where using
ultra-clean materials is most crucial to achieving low backgrounds.

The electroformed copper used in the $\textsc{Majorana Demonstrator}$ was
produced underground at Pacific Northwest National Laboratory and SURF,
and parts were machined and stored underground as well.  This reduces the potential for
cosmogenic activation to produce $^{60}$Co, whose decay can produce photons above the 2039~keV
Q-value \cite{TabRad_v3}, a potential source of background.

$\textsc{Majorana}$ electroforming practices should improve on GERDA
radiopurity for LEGEND-200.  Copper electroforming for LEGEND-200 is underway in the
facilities on the 4850'~level at SURF.  Production of 37~kg of electroformed
copper should be complete by the fall of 2019.

\subsection{Liquid argon veto}
The active liquid argon (LAr) veto employed by GERDA tags external backgrounds
depositing energy in the LAr that subsequently scintillates \cite{Agostini:2017hit}.  
The LAr scintillation light is read out by photomultiplier tubes above and below the detector array,
and a shroud of wavelength-shifting optical fibres surrounding the array read
out by silicon photomultipliers.  An
upgrade this year increased the fibre curtain density and 
added a fibre shroud surrounding the central column of detectors, to improve light collection efficiency.

LEGEND-200 will use a similar design, though different geometries of fibres
between detector columns are currently being studied to optimize light
collection when many more detectors are present in the existing GERDA cryostat.
In addition, light yield and attenuation can be improved with better LAr
purity, as well as doping with Xe, which is currently under investigation.  Using more radiopure fibres
and digitizing the signals in the LAr are also currently under investigation.

The $^{42}$Ar in natural liquid argon is important to consider in the detector
design.  $^{42}$Ar decays to a $^{42}$K ion, which drifts to the charged
detectors, and undergoes $\beta$-decay to $^{42}$Ca with $Q_{\beta}=3.5$~MeV,
providing a source of background to the $0\nu\beta\beta$-decay search. In GERDA
and the future LEGEND-200, the drift of $^{42}$K ions is limited by nylon shrouds around
the detector columns.  In GERDA and LEGEND-200, coincident $\gamma$ 
$^{42}$K-decays are tagged well by the LAr veto, 
and pure $\beta$-decays on the detector surface are cut with $\sim99$\% 
efficiency by pulse shape analysis.

For LEGEND-1000, a design that would remove this background completely makes
use of underground argon free of $^{42}$Ar.  In this configuration, the detectors are
separated into four copper-walled volumes containing underground argon,
submerged in a large natural argon volume.  This design would require 21~tons
(15~m$^{3}$) of underground argon, which is similar to the volume of
underground argon planned to be used by the Darkside-20k experiment \cite{Aalseth:2017fik}.

\subsection{Front-End electronics}
The first stage of the $\textsc{Majorana}$ readout is the Low-Mass Front-End
(LMFE) \cite{Abgrall:2015hna}, located as close as possible to each detector's p+ contact.  Each LMFE contains an MX-11 JFET, and the feedback resistor and capacitor of the resistive
feedback charge sensitive preamplifier.  The feedback loop runs 2.15~m out
through the shielding to the warm preamplifier, and back to the LMFE.  Locating
the first stage of the amplification close to the detector constributes to the
low noise and excellent energy resolution of the $\textsc{Majorana
Demonstrator}$.

Situating the LMFE close to the detectors requires that it have low
backgrounds, through using clean materials and keeping its mass to a minimum.  
To this end, the LMFE features Ti/Au 200/4000~\AA~thick sputtered
traces on a 200~$\mu$m thick amorphous silica substrate, with the feedback
capacitance provided by the capacitance between traces, and a 4000~\AA~sputtered
amorphous germanium feedback resistor. As a result, the LMFE is the most
radiopure front-end ever built. 

The LMFE is therefore in the baseline design for LEGEND-200, where it should
result in reduced backgrounds and noise. Research and development activities
into its performance in liquid argon and its use with longer cables are
currently underway. For LEGEND-1000, research and development into an ASIC
preamplifier that can be placed near the detector is currently underway.

\subsection{Larger $^{\mathrm{enr}}$Ge detectors}
The BEGe-type detectors used by GERDA average 0.66~kg per detector unit, 
and the PPC-type detectors used in the $\textsc{Majorana Demonstrator}$ average 0.85~kg 
per detector unit.  Both of these detector types have excellent energy resolution, 
and superb pulse-shape sensitivity to
reject multi-site and surface background events.  However, larger detectors for
LEGEND would require fewer cables and preamplifier front-ends per active kilogram of detector material,
 which would lower the backgrounds due to these components.  A new detector geometry, the
Inverted-Coaxial Point Contact (ICPC) detector \cite{COOPER201125}, 
allows detectors as large as a normal
coaxial detector, but with similar performance to the BEGes and PPCs employed
by GERDA and the $\textsc{Majorana Demonstrator}$ \cite{Domula:2017mei}.  The baseline mass for these
detectors is 1.5-2.0~kg, with natural germanium detectors near 3~kg also being
studied.  Five enriched ICPC detectors, each around 1.9~kg, have been produced
and deployed in a recent GERDA upgrade, supporting the selection of this detector
 design for LEGEND-200. 

\section{Conclusions}
The next generation of neutrinoless double-beta decay experiments requires
reduced backgrounds and additional mass.  $^{76}$Ge detector arrays have
demonstrated the lowest backgrounds and best energy resolution of all the next
generation experiment technologies.  The LEGEND collaboration plans to take the
best of $\textsc{Majorana}$ and GERDA, and perform additional R\&D to build a
detector with $0\nu\beta\beta$-decay discovery potential at a half-life beyond
$10^{28}$ years.  LEGEND will proceed in a phased approach, starting with the
200~kg LEGEND-200, for which the background reduction goal is
realistic and consistent with the most conservative assay and simulation
results.  Research and development work directed towards \mbox{LEGEND-200} and
LEGEND-1000 is ongoing.

\end{document}